\documentclass[letterpaper]{article} 
\usepackage{aaai2026}  
\usepackage{times}  
\usepackage{helvet}  
\usepackage{courier}  
\usepackage[hyphens]{url}  
\usepackage{graphicx} 
\usepackage{natbib}  
\usepackage{caption} 
\usepackage{algorithm}
\usepackage{algorithmic}
\usepackage{multirow}
\usepackage{amsmath}
\usepackage{multirow}
\usepackage{subcaption}
\usepackage{cleveref}
\usepackage{multicol}
\usepackage{amssymb}
\usepackage{amsfonts}
\usepackage{booktabs}
\usepackage[utf8]{inputenc}
\usepackage[T1]{fontenc}

\newtheorem{Frequency-Domain MLPs}{Definition}

\usepackage{xcolor}
\usepackage{enumitem}
\usepackage[table]{xcolor}
\usepackage{newfloat}
\usepackage{listings}
\DeclareCaptionStyle{ruled}{labelfont=normalfont,labelsep=colon,strut=off} 
\floatstyle{ruled}
\newfloat{listing}{tb}{lst}{}
\floatname{listing}{Listing}

\title{Exploiting Inter-Session Information with Frequency-enhanced \\Dual-Path Networks for Sequential Recommendation}
    
\author{
    Peng He, Yao Liu\thanks{Corresponding Author.}, Yanglei Gan, Run Lin, Tingting Dai, Qiao Liu, Xuexin Li\thanks{Work done as part of an internship program at University of Electronic Science and Technology of China.}
}
\affiliations{

    University of Electronic Science and Technology of China, Chengdu, China\\
    \{hepenglk, yangleigan, runlin, ttdai\}@std.uestc.edu.cn,  \{liuyao, qliu\}@uestc.edu.cn, lxuexin9@gmail.com
    
}

\usepackage{bibentry}

\begin{document}

\maketitle

\begin{abstract}
Sequential recommendation (SR) aims to predict a user's next item preference by modeling historical interaction sequences. Recent advances often integrate frequency-domain modules to compensate for self-attention's low-pass nature by restoring the high-frequency signals critical for personalized recommendations. Nevertheless, existing frequency-aware solutions process each session in isolation and optimize exclusively with time-domain objectives. Consequently, they overlook cross-session spectral dependencies and fail to enforce alignment between predicted and actual spectral signatures, leaving valuable frequency information under-exploited. To this end, we propose \textbf{FreqRec}, a \textbf{Freq}uency-Enhanced Dual-Path Network for sequential \textbf{Rec}ommendation that jointly captures inter-session and intra-session behaviors via a learnable Frequency-domain Multi-layer Perceptron. Moreover, FreqRec is optimized under a composite objective that combines cross entropy with a frequency-domain consistency loss, explicitly aligning predicted and true spectral signatures. Extensive experiments on three benchmarks show that FreqRec surpasses strong baselines and remains robust under data sparsity and noisy-log conditions. Our code is available at: \url{https://github.com/AONE-NLP/FreqRec}.

\end{abstract}

\section{Introduction}
Sequential recommendation (SR) \cite{fang2020deep,wang2019sequential} seeks to infer users' future item preferences by modeling their historical interaction sequences. Such methods have become indispensable in a range of applications, including personalized content delivery \cite{liang2006personalized} and social media feeds \cite{vombatkere2024tiktok}. Over time, these recommenders have undergone significant advances from traditional approaches like Markov chain \cite{rendle2010factorizing} to convolutional
neural networks (CNNs) \cite{tang2018personalized,niu2024diffusion} and recurrent neural networks (RNNs) \cite{donkers2017sequential,liu2018stamp}, and more recently, self-attention mechanisms inspired by the Transformer family. These Transformer-based models \cite{kang2018self,sun2019bert4rec,lin2024multi} offer both improved training efficiency and stronger predictive accuracy. 

Despite these advances, transformer-based approaches are particularly vulnerable to the noisy signals inherent in logged user data \cite{agichtein2006improving}, which can lead models to overfit and degrade generalization \cite{ying2019overview,liu2023distribution}. Moreover, the global receptive field of self-attention acts similarly to a low-pass filter that attenuates high-frequency components \cite{du2023frequency}. Consequently, abrupt shifts or periodic consumption in user behavior can be smoothed out \cite{wang22anti,dovonon2024setting}, erasing fine-grained dynamics that are crucial for personalized recommendations.

\begin{figure}[t]
	\centering
	\subfloat[Sports \& Outdoors]{
		\includegraphics[scale=0.18]{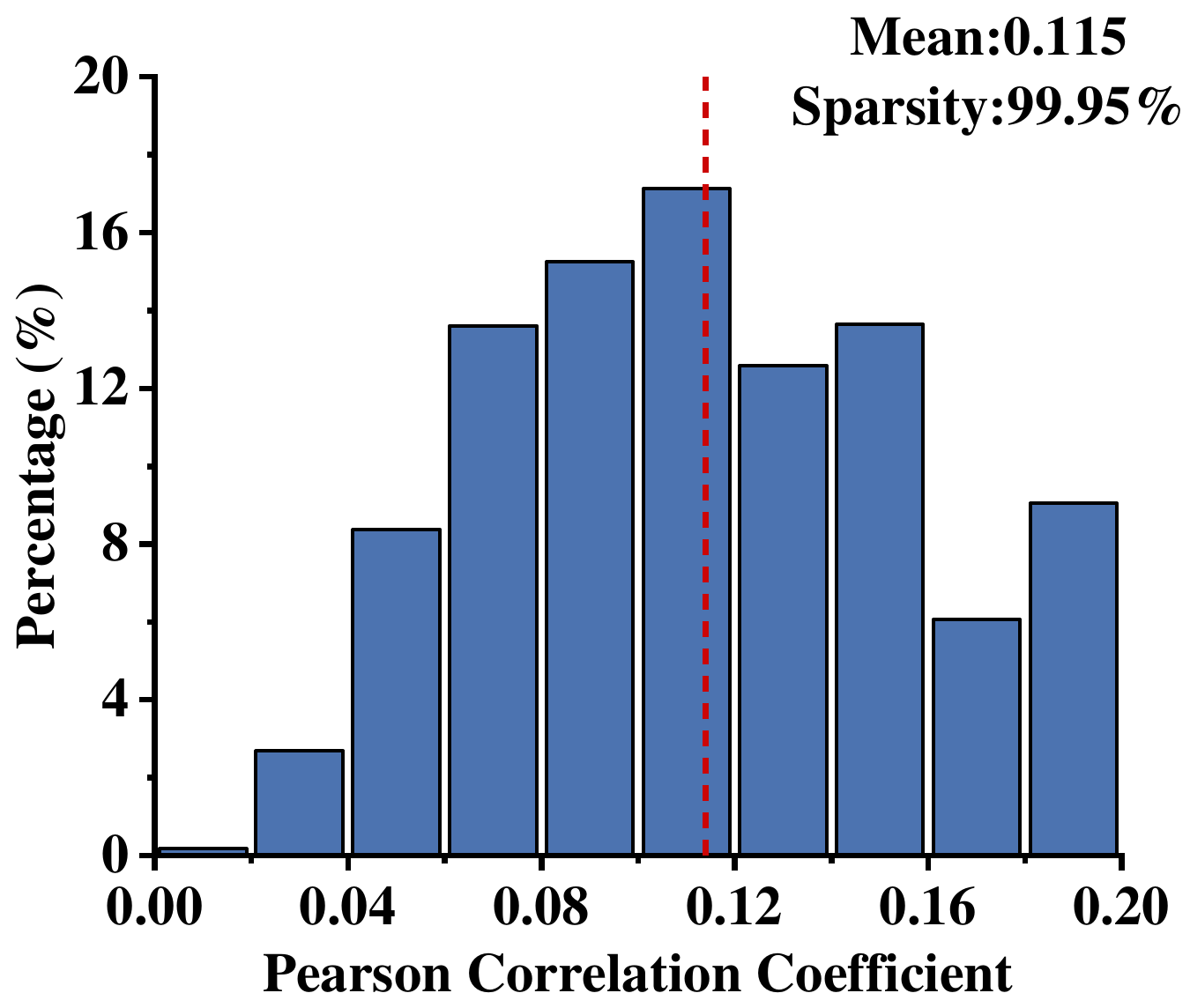}
	}%
	\subfloat[Beauty]{
		\includegraphics[scale=0.18]{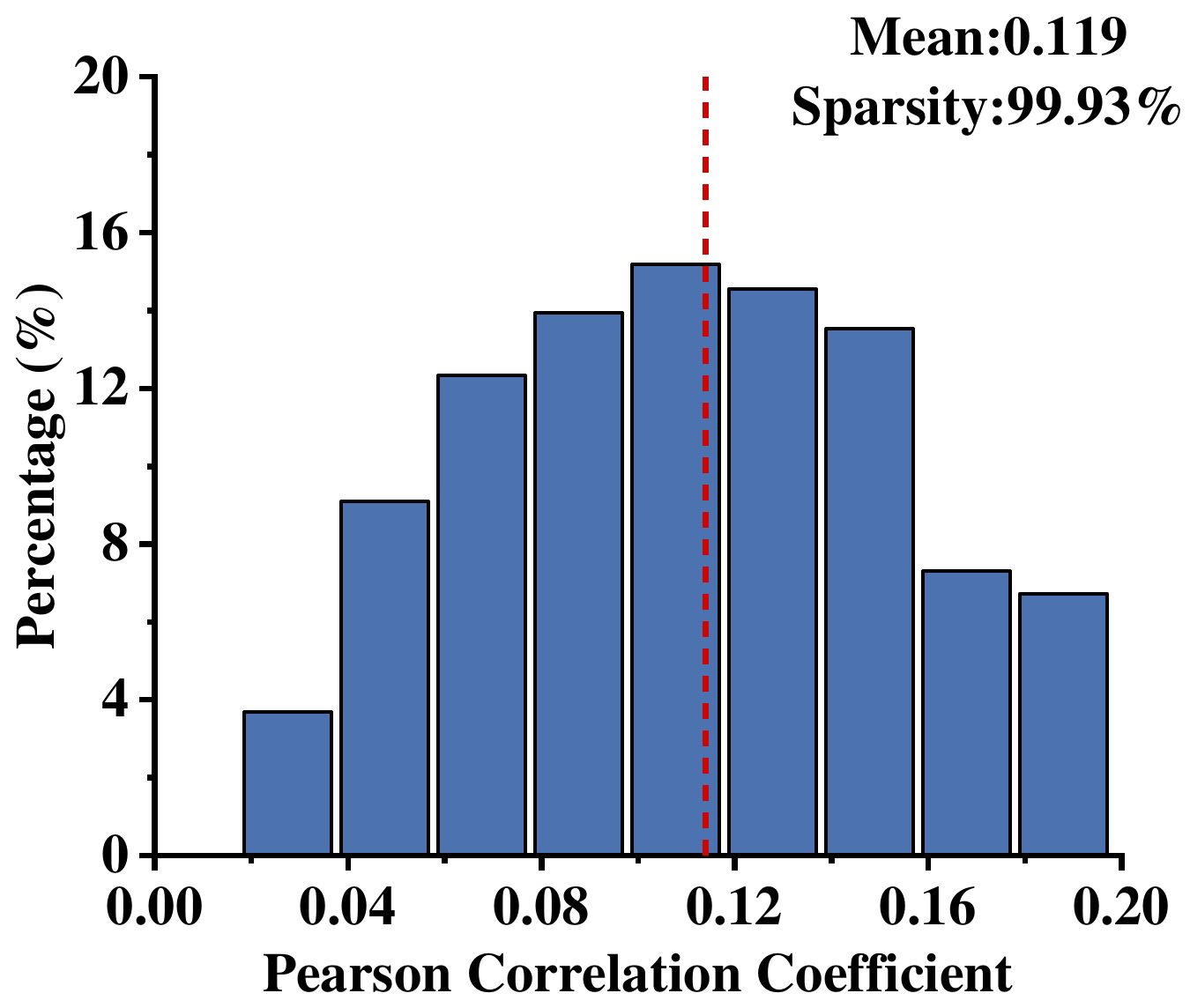}
	}%
	\centering
	\caption{The distribution of Pearson correlation coefficient between sessions sharing at least one item on Sports \& Outdoors and Beauty datasets, respectively.}
	\label{intro}
\end{figure}

Recent work has explored two complementary strategies to address these limitations in the frequency domain. One line of work \cite{zhou2022filter,du2023contrastive,liu2023distribution,kim2025diff} applies explicit spectral filtering to suppress spurious signals and highlight essential sequence features. Another line \cite{du2023frequency,shin2024attentive} integrates the Fourier transform modules alongside self-attention to recover periodic patterns and complementary high-frequency information. Despite their state-of-the-art performance, frequency-based methods introduce new challenges while answering the aforementioned questions: 

\begin{itemize}[leftmargin=*]
    \item \textbf{Neglect of Inter-Session Dependencies.} Contemporary frequency-domain models, such as FMLPRec \cite{du2023frequency}, BSARec \cite{shin2024attentive} treat each session in isolation, forfeiting the rich interaction patterns that span multiple sessions. Given that individual sessions are inherently short \cite{qiu2020exploiting}, ignoring these external signals amplifies data sparsity and undermines predictive accuracy. Figure \ref{intro} shows strong positive correlations between sessions that share at least one item on benchmark datasets, underscoring the importance of inter-session patterns. Graph-based solutions attempt to compensate by constructing inter-session \cite{guo2022evolutionary, qiao2023bi} or global session graphs \cite{wang2020global,wang2023modeling}, but their efficacy depends on meticulously hand-crafted edge rules that are expensive to tune and fragile under sparsity and noise \cite{liu2023enhancing}.
    \item \textbf{Frequency Features Underserved by Time-Domain Objectives.} Most sequential recommendation frameworks optimize losses defined purely in the time domain, such as (Binary) Cross-Entropy \cite{sun2019bert4rec,shin2024attentive}, Contrastive Loss \cite{chen2022intent}, and Bayesian Personalized Ranking Loss \cite{rendle2009bpr,liu2025preference}. Although effective for aligning predicted and observed target items, these objectives provide no direct incentive to capture frequency-domain structure. Consequently, periodic or high-frequency behavioral signals remain under-utilized, limiting the model's capacity to represent the full spectrum of user dynamics.
\end{itemize}

Considering the above issues, we present \textbf{FreqRec}, a \textbf{Freq}uency-Enhanced Dual-Path Network for sequential \textbf{Rec}ommendation that jointly models cohort-level and user-specific behaviors. Specifically, FreqRec encodes the interaction history with self-attention to preserve long-range contextual dependencies, while channels the sequence through two complementary frequency paths: a Global Spectral Aggregator that distills cohort-level rhythms across sessions, and a Local Spectral Refiner that sharpens user-specific nuances. Both paths employ a unified Frequency-Domain Multi-layer Perceptrons (MLPs), whose complex-valued filters are learned end-to-end, eliminating the need for manual frequency cut-offs. The resulting representations are fused in either parallel or serial mode and reconciled with the contextual branch via a gated residual update. To ensure FreqRec learns meaningful spectral features, we augment the standard cross-entropy loss with a frequency-domain consistency loss that explicitly aligns the predicted and ground-truth spectral signatures. Our contribution are three-fold:

\begin{itemize}[leftmargin=*]
    \item We propose a \textbf{Freq}uency-enhanced dual-path network for sequential \textbf{Rec}ommendation (\textbf{FreqRec}), that jointly captures inter-session and intra-session behaviors by integrating a learnable, complex-valued Fourier Transform.
    \item We introduce a frequency-domain consistency loss that explicitly aligns the model's predicted and ground-truth spectral coefficients, enforcing the recovery of high-frequency interaction patterns often smoothed out by standard self-attention mechanisms.
    \item Comprehensive experiments conducted on three real-world benchmarks, demonstrating that FreqRec outperforms state-of-the-art baselines, and remains robust under varying data sparsity and noisy-log scenarios.
\end{itemize}

\section{Related Works}
\subsection{Sequential Recommendation}
Sequential recommendation (SR) aims to learn from users' historical interaction sequences to deliver personalized recommendations \cite{gan2025pareto}. Mainstream SR models predominantly adopt deep learning frameworks: Convolutional Neural Networks (CNNs) \cite{tang2018personalized,niu2024diffusion} and Recurrent Neural Networks (RNNs)  \cite{li2018learning,liu2018stamp} were initially widely employed. For instance, Caser \cite{tang2018personalized} pioneered the treatment of items in recommendation sequences, utilizing convolutional operations to capture item-item interactions. GRU4Rec \cite{Hidasi2015SessionbasedRW} successfully implemented a GRU network to construct a SR framework for modeling long-term dependencies between users-items. Transformers \cite{vaswani2017attention}, owing to their exceptional long-term contextual modeling capabilities and parallel computation efficiency, have gained widespread adoption in SR. Models such as SASRec \cite{kang2018self} and BERT4Rec \cite{sun2019bert4rec} use self-attention to adaptively weight historical interactions, thereby more effectively capturing shifts in user intent. However, the global receptive field of self-attention acts as a low-pass filter, smoothing out abrupt changes in interests in user behavior \cite{du2023frequency}.

To recover these crucial spectral cues, a growing body of work integrates frequency-domain processing within SR models. One line of research designs explicit spectral filters to attenuate noise and amplify informative frequency bands. For instance, Zhou et al. \cite{zhou2022filter} propose a learned band-pass filter that adapts its frequency response to sequence statistics, suppressing spurious interactions while retaining periodic patterns. SLIME4Rec \cite{du2023contrastive} employs contrastive learning on spectral representations to improve robustness against noisy logs, while DIFF \cite{kim2025diff} leverages DFT-based filtering, combined with side information, to remove short-term fluctuations and enhance recommendation accuracy. A second strand embeds Fourier transform operations directly into the network architecture to recover high-frequency dynamics that self-attention may smooth over. FEARec \cite{du2023frequency} and BSARec \cite{shin2024attentive} utilize DFT to capture high-frequency information, counteracting the inherent low-pass filtering properties of self-attention mechanisms. Although these methods achieve strong empirical gains, they process each session in isolation and optimize solely with time-domain losses. These gaps motivate our development of a dual-path architecture that simultaneously models inter-session and intra-session spectral dependencies while incorporating a frequency-domain consistency objective.

\begin{figure*}[t]
  \centering
  \includegraphics[width=0.9\linewidth]{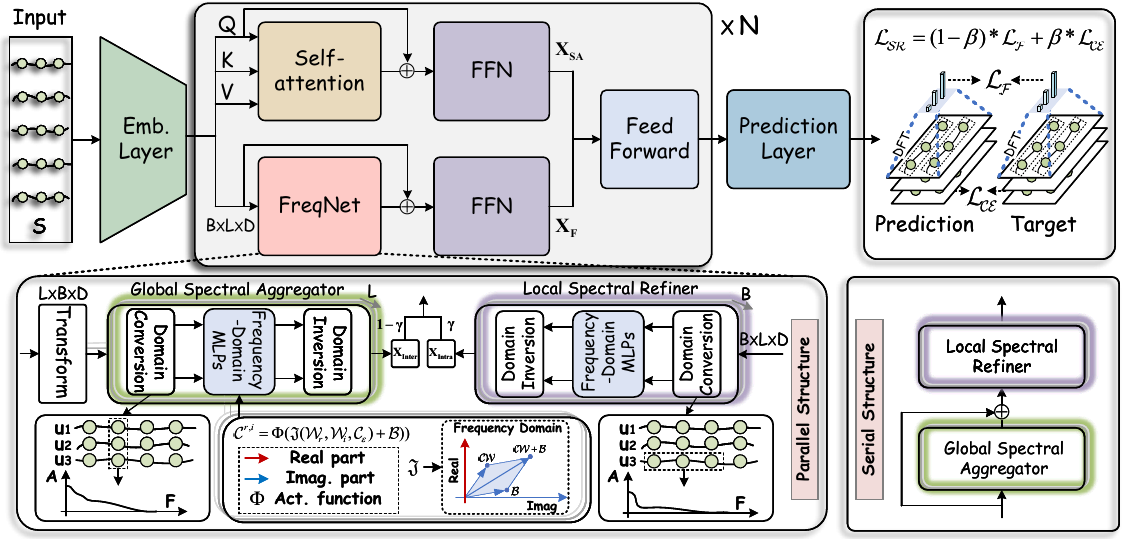}
  \caption{The overall framework of FreqRec. User sequences are embedded and then processed by two parallel paths: a self-attention branch and the \textit{FreqNet} branch. FreqNet contains a \textit{Global Spectral Aggregator} that performs batch-axis DFT $\rightarrow$ FreqMLPs $\rightarrow$ IDFT to distill cohort-level signals, while a \textit{Local Spectral Refiner} applies the same pipeline along the temporal axis to refine user-specific cues. The two spectral modules can be fused in either parallel or serial form. Training minimizes a hybrid objective that couples standard cross-entropy prediction with \textit{frequency-domain consistency loss}.}
  \label{fig:model}
\end{figure*}

\subsection{DFT for Sequential Modeling}
The Discrete Fourier Transform (DFT) is a cornerstone of digital signal processing, decomposing time-domain signals into their constituent frequency components \cite{duhamel1990fast,frigo2005design}. Leveraging its inherent ability to efficiently capture global periodicities and complex patterns, DFT-based modules have recently been adopted to effectively enhance sequential representations. This trend is evident across diverse domains, including time-series forecasting \cite{yi2023frequency,yang2024frequency}, and even natural language processing \cite{tamkin2020language,lee2022fnet}. Building on this trend, our work further extends DFT's applicability by jointly modeling inter- and intra-session spectral dynamics.

\section{Preliminaries}
\noindent\textbf{Sequential Recommendation.} 
Let $\mathcal{U}$ and $\mathcal{I}$ denote the sets of users and items, respectively. For each user $u \in \mathcal{U}$, we observe a chronologically ordered interaction sequence $\mathcal{S}^u = [i_1, i_2, \dots, i_{|\mathcal{S}^u|}]$, where $i_j \in \mathcal{I}$ signifies the $j$-th item which $u$ has interacted and $|\mathcal{S}^u|$ is the sequence length. The goal of SR task is to predict the next item $i^*$ that user $u$ is most probable to consume, given the historical sequence $\mathcal{S}^u$:
\begin{equation}
i_{|\mathcal{S}^u|+1}^{u} = \arg\max_{i^*\in\mathcal{I}}p(i^*|\mathcal{S}^u),
\end{equation} 
where \(p(i^*|\mathcal{S}^u)\) stands for the conditional probability of user $u$ will interact with item $i^*$ next.  

\noindent\textbf{Discrete Fourier Transform (DFT).} 
The Discrete Fourier Transform provides an exact, bijective mapping between a sequence in the time domain and its representation in the frequency domain, thereby enabling precise analysis and manipulation of periodic and noisy components. Let $\mathbf{T} = \left[T_0, T_1, \ldots, T_{L-1}\right] \in \mathbb{R}^L$ denote a real-valued sequence of length $L$. The forward transform, denoted $\mathcal{F}: \mathbb{R}^L \to \mathbb{C}^L$, computes complex-valued coefficients $\mathcal{C}_k$ that encode the amplitude and phase information at each frequency $k$. Its inverse, $\mathcal{F}^{-1}:\mathbb{C}^L \to \mathbb{R}^L$, reconstructs the original sequence exactly from those coefficients.

\begin{itemize}[leftmargin=*]
    \item \textbf{Domain Conversion ($\mathcal{F}$)}. The sequence $T_n$ is mapped to its complex frequency coefficients $\mathcal{C}_k$ via the DFT: 
    \begin{equation}\mathcal{C}_k=\sum_{n=0}^{L-1}T_n\cdot e^{-j2\pi kn/L}, k=0,1, \dots, L-1\end{equation}
    where $n$ and $k$ are the discrete time and frequency indices ($0 \leq n$, $k < L$), and $j$ is the imaginary unit.
    Using Euler's formula, this is separated into its real and imaginary parts:
    \begin{equation}
    \begin{aligned}
    \mathcal{C}_k = \textbf{Real}(\mathcal{C}_k) + j\textbf{Imag}(\mathcal{C}_k), \\
    \textbf{Real}\left(\mathcal{C}_k\right) = \sum_{n=0}^{L-1} T_n \cos\left(\frac{2\pi k n}{L}\right), \\
    \textbf{Imag}\left(\mathcal{C}_k\right) = - \sum_{n=0}^{L-1} T_n \sin\left(\frac{2\pi k n}{L}\right).
    \end{aligned}
    \end{equation}
    \item \textbf{Domain Inversion ($\mathcal{F}^{-1}$)}. We recover the original time-domain sequence $T_n$ via the inverse discrete transform:
    \begin{equation}
    T_n = \frac{1}{L} \sum_{k=0}^{L-1} \mathcal{C}_k \cdot e^{j 2\pi k n / L}
    \end{equation}
\end{itemize}
For brevity, we refer to the forward and inverse mappings simply as $\mathcal{F}$ and $\mathcal{F}^{-1}$.


\section{Methodology}
\subsection{Historical Sequence Encoding}
Suppose each user has a historical interaction sequence $\mathcal{S}^u = [i_1, i_2, \cdots, i_{n-1}]$. We first map $\mathcal{S}^u$ into item embedding $\mathcal{E}_u = \mathcal{M}[\mathcal{S}^u] \in \mathbb{R}^{L\times D}$ via $\mathcal{M} \in \mathbb{R}^{\mathcal{I}\times D}$. We then add a positional embedding $P \in \mathbb{R}^{L\times D}$ and apply layer normalization (LN) to obtain the sequence embedding:
\begin{equation}
    \mathcal{E} = \mathrm{Dropout}(\mathrm{LN}(\mathcal{E}_u+P)).
\end{equation}

\subsection{Dual-Path Representation Learning}
\subsubsection{Contextual-enhanced Representation.}
We apply self-attention to $\mathcal{E}$ to obtain its contextual representation:
\begin{equation}
\begin{aligned}
\mathbf{X_{SA}} = \text{Self-Attention}(\mathcal{E}),\\
\mathbf{X_{SA}} = \mathrm{FFN}(\mathbf{X_{SA}}+\mathcal{E}),
\end{aligned}
\end{equation}
where $\mathcal{E} \in \mathbb{R}^{B \times L \times D}$. FFN denotes Feed Forward Network.

\subsubsection{Frequency-enhanced Network.}\label{FLA}
\begin{figure}[t]
  \centering
  \includegraphics[width=0.75\linewidth]{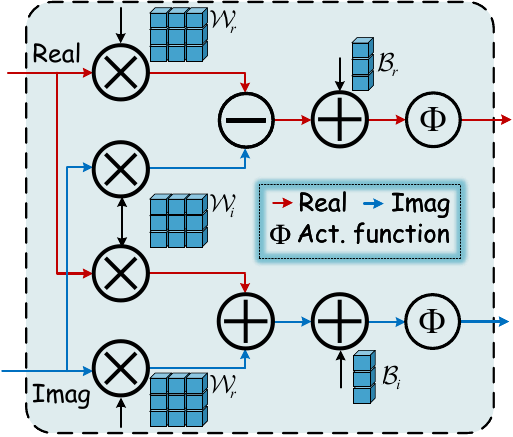}
  \caption{Demonstration of Frequency-Domain MLP.}
  \label{fig:FDMLP}
\end{figure}

As shown in Figure \ref{fig:model}, given the sequence embedding $\mathcal{E}$, we first project it into the frequency domain using the DFT:
\begin{equation}
\begin{aligned}
\centering
\mathcal{C}_{\mathcal{E}}= \mathcal{F}(\mathcal{E}) \rightarrow
\mathcal{C}_{\mathcal{E}} = \textbf{Real}(\mathcal{C}_{\mathcal{E}}) + j\textbf{Imag}(\mathcal{C}_{\mathcal{E}}).
\end{aligned}
\end{equation}
Instead of hand-crafting a low-pass or band-pass filter, we propose a \textbf{Frequency-Domain MLP} (\textbf{FreqMLP}) that learns which frequencies to amplify or suppress. As illustrated in Figure \ref{fig:FDMLP}, FreqMLP jointly processes the real and imaginary channels, enabling cross-component interactions:
\begin{equation}
\begin{gathered}
\mathcal{C}^{r}_{\mathcal{E}} = \varPhi(\mathcal{W}_{r} \cdot \textbf{Real}(\mathcal{C}_{\mathcal{E}}) - \mathcal{W}_{i} \cdot \textbf{Imag}(\mathcal{C}_{\mathcal{E}}) +\mathcal{B}_{r}),\\
\mathcal{C}^{i}_{\mathcal{E}} = \varPhi(\mathcal{W}_{i} \cdot \textbf{Real}(\mathcal{C}_{\mathcal{E}}) + \mathcal{W}_{r} \cdot \textbf{Imag}(\mathcal{C}_{\mathcal{E}}) +\mathcal{B}_{i}),\\
\mathcal{C}^{'}_{\mathcal{E}} = \mathcal{C}^{r}_{\mathcal{E}} + j\mathcal{C}^{i}_{\mathcal{E}},
\end{gathered}
\end{equation}
where $\mathcal{W}_{r}, \mathcal{W}_{i} \in \mathbb{R}^{D \times D}$, $\mathcal{B}_{r}, \mathcal{B}_{i} \in \mathbb{R}^{D}$ and $\varPhi$ is a non-linear activation. We recover a refined time-domain representation via the inverse DFT, denoted as $\mathbf{X_{Freq}}=\mathcal{F}^{-1}(\mathcal{C}^{'}_{\mathcal{E}})$. For brevity, we denote the entire pipeline as:
\begin{equation}
\mathbf{X_{Freq}} = \mathcal{F}^{-1}(\textbf{FreqMLP}(\mathcal{F}(\mathcal{E}))).
\end{equation}

\textbf{\textit{Global Spectral Aggregator}} (GSA).
User logs often share clear population-wide rhythms, that are invaluable when a single user's history is sparse. Hence, we treat a mini-batch of historical embedding $\mathcal{E}_{\mathrm{Inter}} \in \mathbb{R}^{L  \times B \times D}$ as a signal over the batch axis, perform a batch-axis DFT, pass the complex coefficients through a dedicated \textbf{FreqMLP} to suppress high-frequency noise and boost low-frequency cohort trends:
\begin{equation}
\mathbf{X_{Inter}} = \mathcal{F}^{-1}(\textbf{FreqMLP}(\mathcal{F}(\mathcal{E}_{\mathrm{Inter}}))).
\end{equation}

\textbf{\textit{Local Spectral Refiner}} (LSR).
Complementing the GSA stage, the LSR captures each user's fine-grained and often transient interests by applying the same frequency filtering along the temporal dimension $L$, given the sequence embedding of users $\mathcal{E}_{\mathrm{Intra}} \in \mathbb{R}^{B\times L \times D}$:
\begin{equation}
\mathbf{X_{Intra}} = \mathcal{F}^{-1}(\textbf{FreqMLP}(\mathcal{F}(\mathcal{E}_{\mathrm{Intra}}))),
\end{equation}
where the input $\mathcal{E}_{\mathrm{Intra}}$ depends on the fusion strategy:
\begin{equation}
\mathcal{E}_{\mathrm{Intra}}= \begin{cases}\mathcal{E}, & \textit{Parallel Fusion} \\ \mathcal{E}+\mathbf{X_{Inter}}. & \textit{Serial Fusion}\end{cases}
\end{equation}

\textbf{\textit{Fusion and Output.}} 
Once we have both the cohort representation $\mathbf{X_{Inter}}$ and the personalized representation $\mathbf{X_{Intra}}$, we integrate them into a single feature map, which can proceed in the following way:
\begin{itemize}[leftmargin=*]
    \item \textit{Parallel Fusion} processes inter- and intra-session streams independently and combines their respective outputs with a weighted sum, which preserves the integrity of the raw sequence features and allows each module to specialize without mutual interference:
\begin{equation}
\mathbf{X}_{\mathrm{F}}=\mathrm{FFN}\left((1-\gamma) * \mathbf{X_{Inter}}+\gamma * \mathbf{X_{Intra}}, \mathcal{E}\right).
\end{equation}
    \item \textit{Serial Fusion} feeds $\mathbf{X_{Inter}}$ directly into the LSR stage, ensuring that individual preference modeling is directly informed by cohort-level patterns:
\begin{equation}
    \mathbf{X}_{\mathrm{F}}=\mathrm{FFN}(\mathbf{X_{Intra}},\mathcal{E}).
\end{equation}
\end{itemize}

Finally, we integrate the contextual-enhanced representation via a gated residual update:
\begin{equation}
\begin{gathered}
    \mathbf{X_{out}}=(1-\alpha)*\mathbf{X_{SA}}+\alpha*\mathbf{X_{F}},\\
    \mathbf{X_{out}}=\mathrm{LN}(\mathrm{Dropout}(\mathrm{GELU}(\mathbf{X_{out}}))+\mathbf{X_{out}}),
\end{gathered}
\end{equation}
where $\mathbf{X_{out}}\in \mathbb{R}^{B \times L \times D}$, and $\alpha$ is the hyper-parameter. The final recommendation scores are obtained by computing the similarity between $\mathbf{X_{out}}$ and the item embedding matrix $\mathcal{M}$:
\begin{equation}
    \bar{y_{i}} = \mathbf{X_{out}}*\mathcal{M}^\top. 
\end{equation}

\begin{table*}[t]
\centering
\fontsize{7.5pt}{8pt} \selectfont 
\renewcommand{\arraystretch}{0.7} 
\setlength{\tabcolsep}{3pt} 
\begin{tabular}{c|cccc|cccc|cccc}
\toprule
\textbf{Dataset} & \multicolumn{4}{c|}{\textbf{Beauty}} & \multicolumn{4}{c|}{\textbf{Sports \& Outdoors}} & \multicolumn{4}{c}{\textbf{Toys \& Games}} \\ \midrule
Metric & HR@10 & HR@20 & NDCG@10 & NDCG@20 & HR@10 & HR@20 & NDCG@10 & NDCG@20 & HR@10 & HR@20 & NDCG@10 & NDCG@20 \\ \midrule
GRU4Rec \citeyearpar{Hidasi2015SessionbasedRW} & 0.0304 & 0.0527 & 0.0147 & 0.0203 & 0.0187 & 0.0303 & 0.0101 & 0.0131 & 0.0211 & 0.0348 & 0.0106 & 0.0140 \\
Caser \citeyearpar{tang2018personalized} & 0.0225 & 0.0403 & 0.0108 & 0.0153 & 0.0163 & 0.0260 & 0.0080 & 0.0104 & 0.0161 & 0.0268 & 0.0079 & 0.0106 \\
SASRec \citeyearpar{kang2018self} & 0.0531 & 0.0823 & 0.0283 & 0.0356 & 0.0298 & 0.0459 & 0.0159 & 0.0200 & 0.0652 & 0.0929 & 0.0366 & 0.0435 \\
BERT4Rec \citeyearpar{sun2019bert4rec}  & 0.0705 & 0.1073 & 0.0387 & 0.0480 & 0.0428 & 0.0649 & 0.0229 & 0.0284 & 0.0635 & 0.0939 & 0.0353 & 0.0430 \\
DuoRec \citeyearpar{xie2022decoupled} & 0.0800 & 0.1088 & 0.0483 & 0.0555 & 0.0446 & 0.0640 & 0.0263 & 0.0312 & 0.0816 & 0.1080 & 0.0515 & 0.0582 \\
ICLRec \citeyearpar{chen2022intent} & 0.0726 & 0.1055 & 0.0400 & 0.0483 & 0.0422 & 0.0632 & 0.0227 & 0.0280 & 0.0826 & 0.1137 & 0.0473 & 0.0552 \\
CL4SRec \citeyearpar{xie2022contrastive} & 0.0706 & 0.0990 & 0.0417 & 0.0488 & 0.0375 & 0.0575 & 0.0201 & 0.0251 & 0.0863 & 0.1143 & 0.0522 & 0.0592 \\
MAERec \citeyearpar{ye2023graph} & 0.0789 & 0.1094 & 0.0472 & 0.0548 & 0.0435 & 0.0645 & 0.0239 & 0.0292 & 0.0823 & 0.1108 & 0.0499 & 0.0570 \\
$\text{SASRec}_\text{F}$$^\spadesuit$ \citeyearpar{lin2024multi}& 0.0849 & 0.1187 & 0.0421 & 0.0506 & 0.0488 & 0.0711 & 0.0231 & 0.0287 & 0.0875 & 0.1082 & 0.0571 & 0.0623 \\
MSSR$^\spadesuit$ \citeyearpar{lin2024multi} & 0.0897 & 0.1281 & 0.0448 & 0.0545 & 0.0540 & 0.0785 & 0.0255 & 0.0317 & 0.0982 & 0.1201 & 0.0608 & 0.0684 \\ \midrule
FMLPRec \citeyearpar{zhou2022filter} & 0.0559 & 0.0869 & 0.0291 & 0.0369 & 0.0336 & 0.0525 & 0.0183 & 0.0231 & 0.0671 & 0.0974 & 0.0365 & 0.0441 \\
FEARec$^\spadesuit$ \citeyearpar{du2023frequency} & 0.0927 & 0.1282 & 0.0555 & 0.0645 & 0.0539 & 0.0783 & 0.0315 & 0.0376 & 0.1005 & 0.1357 & 0.0606 & 0.0708 \\
BSARec$^\spadesuit$ \citeyearpar{shin2024attentive} & 0.0944 & 0.1295 & 0.0574 & 0.0672 & \underline{0.0581} & \underline{0.0830} & 0.0333 & 0.0387 & 0.1023 & 0.1379 & 0.0610 & 0.0713 \\
DIFF$^\spadesuit$ \citeyearpar{kim2025diff} & 0.0940 & \underline{0.1301} & 0.0526 & 0.0631 & 0.0580 & 0.0829 & 0.0310 & 0.0381 & 0.0969 & 0.1210 & 0.0607 & 0.0703 \\ \midrule
\rowcolor{gray!30}
$\text{FreqRec (S)}^*$ & \underline{0.0962} & 0.1299 & \underline{0.0589} & \underline{0.0674} & 0.0571 & 0.0825 & \underline{ 0.0334} & \underline{ 0.0398} & \textbf{0.1046} & \underline{0.1459} & \textbf{0.0655} & \underline{0.0734} \\
\rowcolor{gray!30} 
$\text{FreqRec (P)}^*$ & \textbf{0.0989} & \textbf{0.1359} & \textbf{0.0601} & \textbf{0.0686} & \textbf{0.0583} & \textbf{0.0859} & \textbf{0.0342} & \textbf{0.0401} & \underline{0.1044} & \textbf{0.1468} & \underline{0.0653} & \textbf{0.0735} \\ \midrule
\textit{Improve.} & 4.77\% & 4.46\% & 4.70\% & 2.08\% & 0.34\% & 3.49\% & 2.70\% & 3.62\% & 2.25\% & 6.45\% & 7.38\% & 3.09\% \\ \bottomrule
\end{tabular}
\caption{Overall performance comparison on the three datasets. S and P denote FreqRec's serial and parallel structures, respectively. Best and suboptimal results are \textbf{bold} and \underline{underlined}, respectively. $*$ denotes a statistically significant improvement of FreqRec over the best competing model ($p < 0.05$). Models marked with $^\spadesuit$ indicate values obtained from our own re-implementation of the corresponding baseline, other results retrieved from BSARec \cite{shin2024attentive}. \textit{Improve.} is FreqRec's relative improvement over the strongest baseline. We employ \textbf{FreqRec (P)} as our standard model for the following experiments.} 
\label{overper}
\end{table*}

\subsection{Training Objective }
Our training objective is composed of two parts: the Cross-Entropy (CE) and the Frequency Domain consistency loss.
\subsubsection{Cross-Entropy Loss. }
In the experiment, we select CE loss as the first train loss $\mathcal{L_{CE}}$. CE loss regards predicting future items as a categorization task across the item set. The process of CE loss is expressed as:
\begin{equation}\begin{gathered}
    \mathcal{L_{CE}}=-\mathrm{log}\frac{\mathrm{exp}(\bar{y_{t}})}{\sum_{i\in|\mathcal{I}|}\mathrm{exp}(\bar{y_{i}})},
\end{gathered}\end{equation}
where $\bar{y_{t}}$ is the ground truth item.
\subsubsection{Frequency Domain Loss. }
We transform the prediction $P$ and target $T$ into the frequency domain. 
We define $\mathcal{L}_{\mathrm{L1}}(P,T)=\frac{1}{N}\sum_{i=1}^{N}|P_{i}-T_{i}|$, $\mathcal{L}_{\mathrm{L2}}(P,T)=\frac{1}{N}\sum_{i=1}^{N}(P_{i}-T_{i})^{2}$, and $\mathcal{L}_{\mathrm{mix}}=\frac{1}{2}(\mathcal{L}_{\mathrm{L1}}+\mathcal{L}_{\mathrm{L2}})$. 
The distance function $\mathcal{F_{D}}$ is selected from:
\begin{equation}
\mathcal{F_{D}}(P,T) \in \{\mathcal{L}_{\mathrm{L1}}(P,T), \mathcal{L}_{\mathrm{L2}}(P,T), \mathcal{L}_{\mathrm{mix}}(P,T)\}
\end{equation}
where $P, T \in \mathbb{R}^{B \times L \times D}$. The final frequency-domain loss $\mathcal{L_F}$ separately computes the distance between the real and imaginary parts of $P$ and $T$ after the DFT $\mathcal{F}$:
\begin{equation}\begin{aligned}
\mathcal{L_F}=\mathcal{F_{D}}(\textbf{Real}(\mathcal{F}(P)),\textbf{Real}(\mathcal{F}(T))) \\
+\mathcal{F_{D}}(\textbf{Imag}(\mathcal{F}(P)),\textbf{Imag}(\mathcal{F}(T))).\end{aligned}\end{equation}
Finally, our final loss function $\mathcal{L_{SR}}$ is expressed as:
\begin{equation}
    \mathcal{L_{SR}}=(1-\beta)*\mathcal{L_{F}}+\beta*\mathcal{L_{CE}},
\end{equation}
where $\beta$ control the strengths of the frequency-domain loss $\mathcal{L_{F}}$ and Cross-Entropy loss $\mathcal{L_{CE}}$.

\section{Experiments}
\subsection{Experimental Setup }
\subsubsection{Datasets. }
We conducted a comprehensive evaluation on three widely adopted SR benchmarks\footnote{https://cseweb.ucsd.edu/~jmcauley/datasets/amazon\_v2/}: {\tt Beauty}, {\tt Sports \& Outdoors}, and {\tt Toys \& Games}. These datasets, derived from the Amazon e-commerce platform, contain product reviews from 1996 to 2014 and are widely used for the SR task. We adopted the data preprocessing approach from previous work \cite{shin2024attentive, kim2025diff}. 

\subsubsection{Evaluation Metrics. } 
To comprehensively evaluate the model's recommendation accuracy, we employ the top-$K$ evaluation method commonly used in SR, including Hit Rate (HR@$K$) and Normalized Discounted Cumulative Gain (NDCG@$K$), with $K$ set at 10 and 20. 

\subsubsection{Baselines. }
To comprehensively evaluate FreqRec, we benchmark its performance against fourteen SR baselines:
\begin{itemize}[leftmargin=*]
\item RNN or Transformer-based approaches: GRU4Rec \cite{Hidasi2015SessionbasedRW}, Caser \cite{tang2018personalized}, SASRec \cite{kang2018self}, BERT4Rec \cite{sun2019bert4rec}, DuoRec \cite{qiu2022contrastive}, ICLRec \cite{chen2022intent}, CL4SRec \cite{xie2022contrastive}, MAERec \cite{ye2023graph}, $\text{SASRec}_\text{F}$ \cite{lin2024multi}, MSSR \cite{lin2024multi}.

\item Frequency-based approaches: FMLPRec \cite{zhou2022filter}, FEARec \cite{du2023frequency}, BSARec \cite{shin2024attentive}, DIFF \cite{kim2025diff}.
\end{itemize} 

\subsubsection{Implementation Details}
Experiments were conducted on an NVIDIA RTX 4090 (24 GB) using PyTorch. We employed the Adam optimizer with 10-round early stopping, and a learning rate grid-searched from $\{\text{5e-4}, \text{1e-4}, \text{1e-3}\}$. For fair comparison, we followed prior work \cite{shin2024attentive,kim2025diff} setting $D=64$ and $L=50$. For FreqRec, we grid-searched $\gamma, \alpha \in \{0.1, 0.3, 0.5, 0.7, 0.9\}$, $\beta \in \{0.1, 0.2, 0.3, 0.4, 0.5, 0.6, 0.7, 0.8, 0.9\}$, batch size $B \in \{32, 64, 128, 256, 512\}$, and $\mathcal{F_D} \in \{\mathcal{L}_{\mathrm{L1}}, \mathcal{L}_{\mathrm{L2}}, \mathcal{L}_{\mathrm{mix}}\}$.

\subsection{Overall Performance}
Table \ref{overper} compares FreqRec against leading baselines on {\tt Beauty}, {\tt Sports \& Outdoors}, and {\tt Toys \& Games} benchmarks. FreqRec consistently delivers superior performance in terms of HR@10/20 and NDCG@10/20. A closer examination yields the following key insights:
\begin{itemize}[leftmargin=*]
    \item \textbf{Frequency-based approaches demonstrate strong performance.} Consistent with prior findings \cite{shin2024attentive}, methods leveraging frequency-based representations outperform transformer-based models in most cases. On {\tt Beauty}, HR@10/20 for BSARec is 0.0944/0.1295 compared to 0.0897/0.1281 for MSSR. This gain arises because frequency-domain modules recover high-frequency and periodic signals, patterns that standard self-attention, acting as a low-pass filter, inherently attenuates.
    \item \textbf{FreqRec exhibits superior performance over Frequency-based models.} Across all three benchmark datasets, FreqRec demonstrates consistent improvements over the strongest baseline BSARec, with performance gains ranging from 0.34\% to 7.38\%. On the {\tt Toys \& Games} dataset, FreqRec achieves an NDCG@10 of 0.0655, while BSARec attains 0.0610, representing a performance improvement of 7.38\%. These results provide compelling evidence for the effectiveness of the learnable inter/intra-session frequency MLPs and Fourier loss mechanism in enabling accurate recommendations.
    \item \textbf{Parallel modeling of inter/intra sequence interactions outperforms serial approaches.} We implement two FreqRec variants, one that captures inter- and intra-session dependencies in parallel, and another that processes them sequentially. The parallel design consistently achieves superior performance. We attribute this gap to an information bottleneck in the serial design, where the user-perspective features overwrite raw sequence signals, impeding the recovery of fine-grained dynamics. In contrast, Parallel Fusion preserves the integrity of both spectral streams, enabling the model to leverage complementary cohort and individual cues without mutual interference.

\end{itemize}

\begin{table}[t]
\centering
\fontsize{7.3pt}{8pt} \selectfont 
\renewcommand{\arraystretch}{0.8} 
\setlength{\tabcolsep}{1.8pt} 
\begin{tabular}{@{}c|cccc|cccc@{}}
\toprule
& \multicolumn{4}{c|}{\textbf{Beauty}} & \multicolumn{4}{c}{\textbf{Toys \& Games}} \\ \cmidrule(l){2-9} 
\multirow{-2}{*}{\textbf{Model}} & H@10 & H@20 & N@10 & N@20 & H@10 & H@20 & N@10 & N@20 \\ \midrule
w/o SA & 0.0959 & 0.1279 & 0.0587 & 0.0667 & 0.1034 & 0.1338 & 0.0644 & 0.0713 \\
w/o GSA & 0.0881 & 0.1202 & 0.0537 & 0.0618 & 0.0954 & 0.1295 & 0.0568 & 0.0672 \\
w/o LSR & 0.0888 & 0.1215 & 0.0533 & 0.0616 & 0.0961 & 0.1268 & 0.0606 & 0.0684 \\
w/o GSA+LSR & 0.0787 & 0.1097 & 0.0481 & 0.0559 & 0.0719 & 0.0956 & 0.0436 & 0.0495 \\ \midrule
w/o $\mathcal{L_F}$& 0.0969 & 0.1300 & 0.0582 & 0.0665 & 0.0990 & 0.1342 & 0.0619 & 0.0708 \\
w/o $\mathcal{L_{CE}}$ & 0.0807 & 0.1081 & 0.0477 & 0.0548 & 0.0714 & 0.0975 & 0.0434 & 0.0499 \\ \midrule
FreqRec & \textbf{0.0989} & \textbf{0.1359} & \textbf{0.0601} & \textbf{0.0686} & \textbf{0.1044} & \textbf{0.1468} & \textbf{0.0653} & \textbf{0.0735} \\ \bottomrule
\end{tabular}
\caption{Ablation study on {\tt Beauty} and {\tt Toys \& Games}. "H" and "N" denote HR@K and NDCG@K, respectively.}
\label{abla}
\end{table}
\subsection{Ablation Studies}\label{sec:ablation}
To illustrate the effectiveness of our proposed Frequency domain MLPs and the Fourier loss in FreqRec, we conduct ablation studies on {\tt Beauty} and {\tt Toys \& Games} datasets. As demonstrated in Table \ref{abla}, when we disable inter-session interaction modeling (w/o $\mathrm{GSA.}$) or intra-session interaction modeling (w/o $\mathrm{LSR.}$), overall performance degrades by a comparable margin. Crucially, omitting both components results in over 20\% on average performance drop relative to FreqRec, confirming that both inter and intra frequency signals complement self-attention representations. Furthermore, the slightly greater performance drop observed for the w/o $\mathrm{GSA.}$ variant highlights the pivotal role of capturing mutual interests between sequences/users in enhancing recommendation accuracy. Beyond architectural components, the auxiliary frequency-domain loss $\mathcal{L_{F}}$ also proves essential. Disabling this loss produces a noticeable performance decline on both datasets: on {\tt Toys \& Games}, HR@20 drops from 0.1468 to 0.1342, and NDCG@20 from 0.0735 to 0.0708. Conversely, without cross-entropy loss $\mathcal{L_{CE}}$ yields the most severe degradation, reaffirming that point-wise prediction accuracy remains the primary supervisory signal. Taken together, these results demonstrate that FreqRec's performance gains stem from the synergy between the joint optimization of cross-entropy and frequency-domain loss.

\begin{table}[t]
\tiny
\centering
\fontsize{7.3pt}{8pt} \selectfont 
\setlength{\tabcolsep}{5pt} 
\renewcommand{\arraystretch}{0.6} 
\begin{tabular}{@{}c|cc|cc@{}}
\toprule
 & \multicolumn{2}{c|}{\textbf{Beauty}} & \multicolumn{2}{c}{\textbf{Toys \& Games}} \\ \cmidrule(l){2-5} 
\multirow{-2}{*}{Model} & HR@10& NDCG@10 & HR@10 & NDCG@10 \\ \midrule
SASRec & 0.0531 & 0.0283 & 0.0640 & 0.0366 \\
$+\mathcal{L_F}$ & 0.0592$^{\textcolor{red}{\text{+11.49\%}}}$ & 0.0309$^{\textcolor{red}{\text{+9.19\%}}}$ & 0.0667$^{\textcolor{red}{\text{+4.22\%}}}$ & 0.0374$^{\textcolor{red}{\text{+2.19\%}}}$ \\ \midrule
DuoRec & 0.0890 & 0.0542 & 0.1027 & 0.0635 \\
$+\mathcal{L_F}$ & 0.0911$^{\textcolor{red}{\text{+2.36\%}}}$ & 0.0554$^{\textcolor{red}{\text{+2.21\%}}}$ & 0.1030$^{\textcolor{red}{\text{+0.29\%}}}$ & 0.0647$^{\textcolor{red}{\text{+1.89\%}}}$ \\ \midrule
FMLPRec & 0.0559 & 0.0291 & 0.0628 & 0.0356 \\
$+\mathcal{L_F}$& 0.0644$^{\textcolor{red}{\text{+15.21\%}}}$  & 0.0349$^{\textcolor{red}{\text{+19.93\%} }}$ & 0.0686$^{\textcolor{red}{\text{+9.24\%}}}$ & 0.0389$^{\textcolor{red}{\text{+9.27\%}}}$  \\ \midrule
BSARec & 0.0871 & 0.0437 & 0.1065 & 0.0658 \\
$+\mathcal{L_F}$ & 0.0972$^{\textcolor{red}{\text{+11.60\%}}}$  & 0.0640$^{\textcolor{red}{\text{+46.45\%}}}$  & 0.1077$^{\textcolor{red}{\text{+1.13\%}}}$  & 0.0668$^{\textcolor{red}{\text{+1.52\%}}}$  \\ \midrule
\textit{\textbf{AVG. Improve}} & \textbf{10.165\%} & \textbf{19.455\%} & \textbf{3.720\%} & \textbf{3.718\%} \\ \bottomrule
\end{tabular}
\caption{Impact of the frequency-domain consistency loss ($+\mathcal{L_F}$). \textcolor{red}{Red} superscripts indicate the improvement rates.}
\label{FFTloss}
\end{table}

\begin{table}
\tiny
\centering
\fontsize{8pt}{8pt} \selectfont 
\renewcommand{\arraystretch}{0.5} 
\setlength{\tabcolsep}{4.8pt} 
\begin{tabular}{@{}c|cc|cc|cc@{}}
\toprule
 & \multicolumn{2}{c|}{\textbf{Automotive}} & \multicolumn{2}{c|}{\textbf{CDs}} & \multicolumn{2}{c}{\textbf{Grocery}} \\ \cmidrule(l){2-7} 
\multirow{-2}{*}{Model} & H@10 & N@10 &  H@10 & N@10 & H@10 & N@10 \\ \midrule
FMLPRec & 0.0320 & 0.0204 & 0.0367 & 0.0207 & 0.1043 & 0.0705 \\
BSARec & 0.0514 & 0.0430 & 0.0537 & 0.0438 & 0.1414 & 0.1215 \\ \midrule
w/o GSA & 0.0495 & 0.0411 & 0.0520 & 0.0425 & 0.1371 & 0.1177 \\
w/o LSR & 0.0507 & 0.0423 & 0.0524 & 0.0443 & 0.1379 & 0.1182 \\
w/o GSA+LSR & 0.0463 & 0.0387 & 0.0518 & 0.0402 & 0.1346 & 0.1149 \\
w/o $\mathcal{L_F}$ & 0.0501 & 0.0426 & 0.0538 & 0.0426 & 0.1402 & 0.1201 \\ \midrule
FreqRec & \textbf{0.0521} & \textbf{0.0445} & \textbf{0.0549} & \textbf{0.0475} & \textbf{0.1434} & \textbf{0.1242} \\ \bottomrule
\end{tabular}
\caption{Performance under noisy scenario in terms of HR@10 and NDCG@10 across the three target domains.}
\label{cross_domain}
\end{table}

\subsection{Effectiveness of Frequency Domain Loss}
\noindent In this section, we assess the effectiveness and flexibility of our Fourier loss $\mathcal{L_F}$ by integrating it into four representative sequential recommendation baselines on the {\tt Beauty} and {\tt Toys \& Games} datasets. To ensure experimental fairness, we adhere strictly to each model's original implementation and hyper-parameter settings. As shown in Table \ref{FFTloss}, our proposed Fourier loss consistently improves baseline performance in a true plug-and-play fashion, recording average improvements over both datasets in terms of HR@10 and NDCG@10 ranging from 3.70\% to 19.45\%. It is noteworthy that models which already exploit frequency domain features (FMLPRec and BSARec) derive the greatest benefit. FMLPRec's HR@10 climbs by 15.21\% (from 0.0559 to 0.0644) and NDCG@10 by 19.93\% (from 0.0291 to 0.0349), while BSARec secures a 46.45\% gain in NDCG@10 (from 0.0437 to 0.0639). These considerable improvements underscore that explicitly penalizing frequency discrepancies provides a complementary signal to cross-entropy loss, guiding DFT-based models toward more discriminative feature learning. 

\begin{figure*}[t]
	\centering
	\subfloat[Impact of $\gamma$ on Beauty]{
		\includegraphics[scale=0.25]{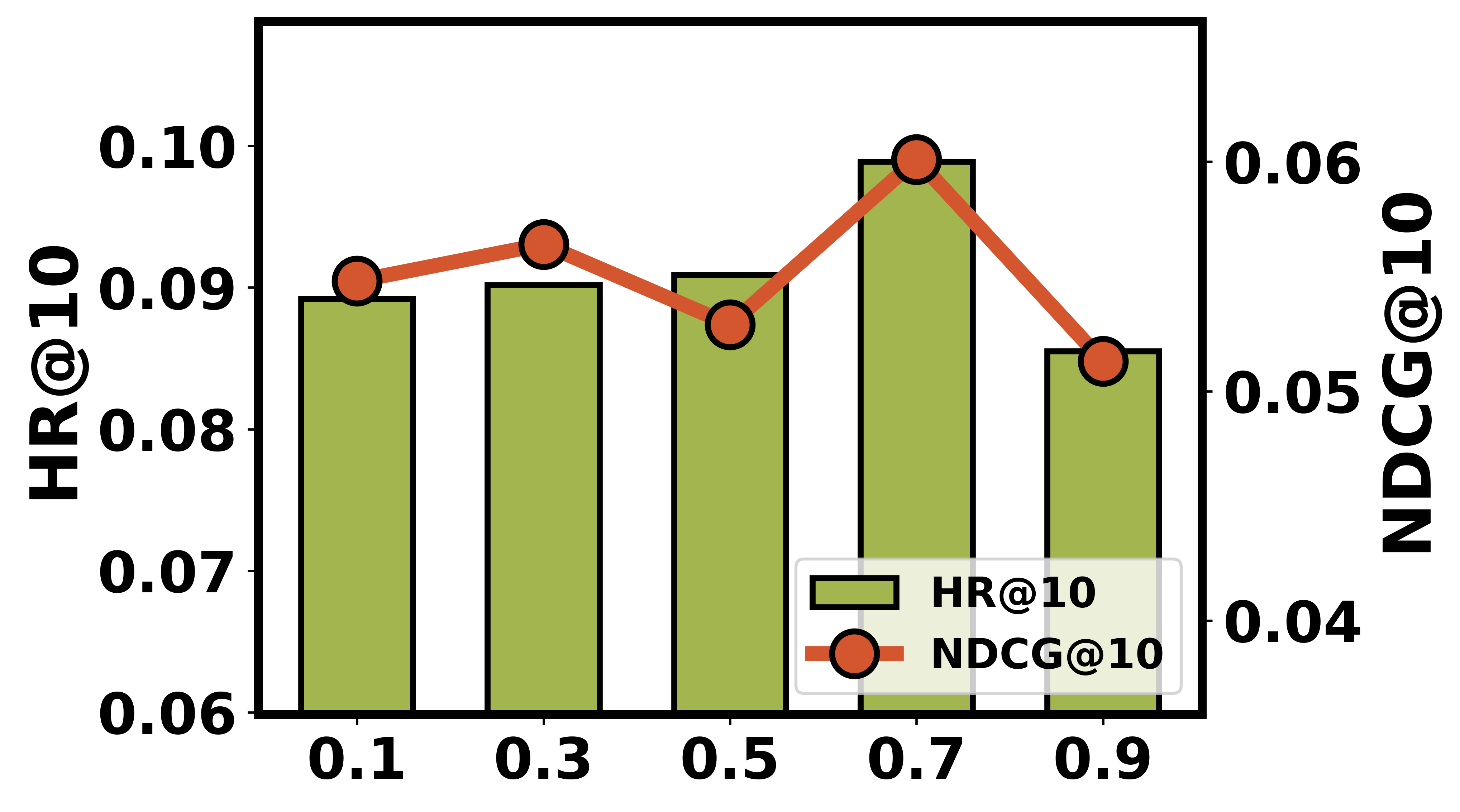}
	}%
	\subfloat[Impact of $\alpha$ on Beauty]{
		\includegraphics[scale=0.25]{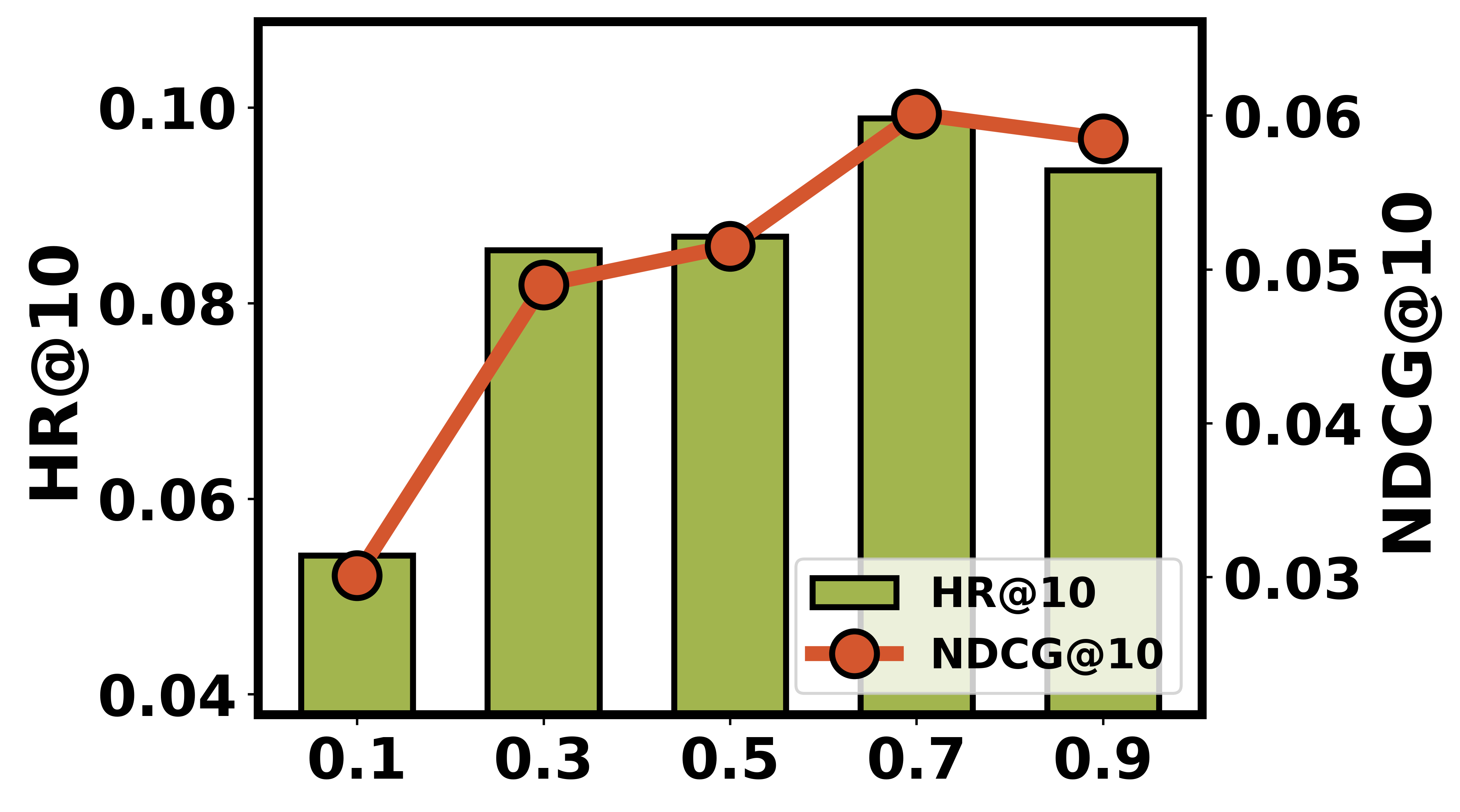}
	}%
 	\subfloat[Impact of $\beta$ on Beauty]{
		\includegraphics[scale=0.25]{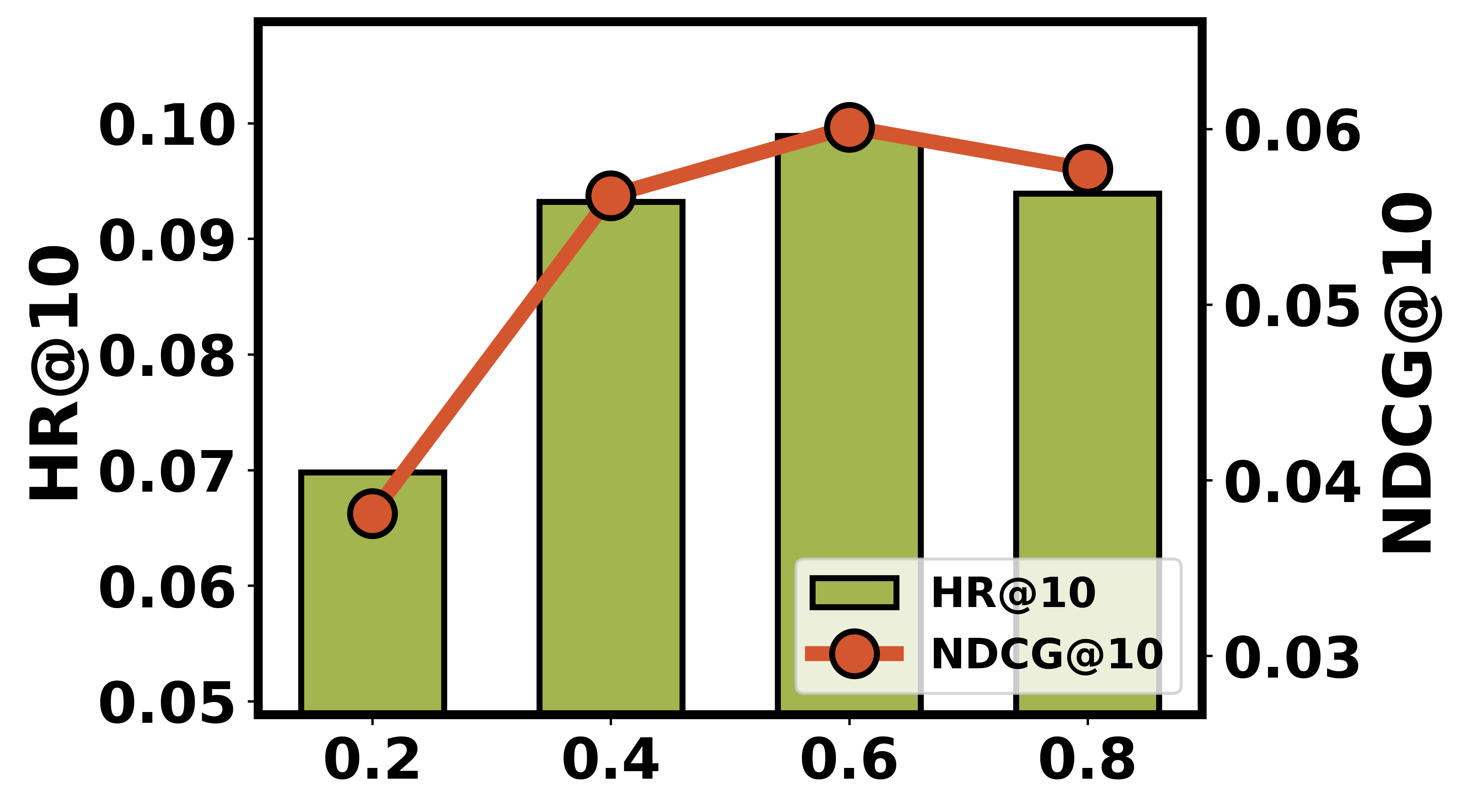}
	}%
 	\subfloat[Impact of $B$ on Beauty]{
		\includegraphics[scale=0.25]{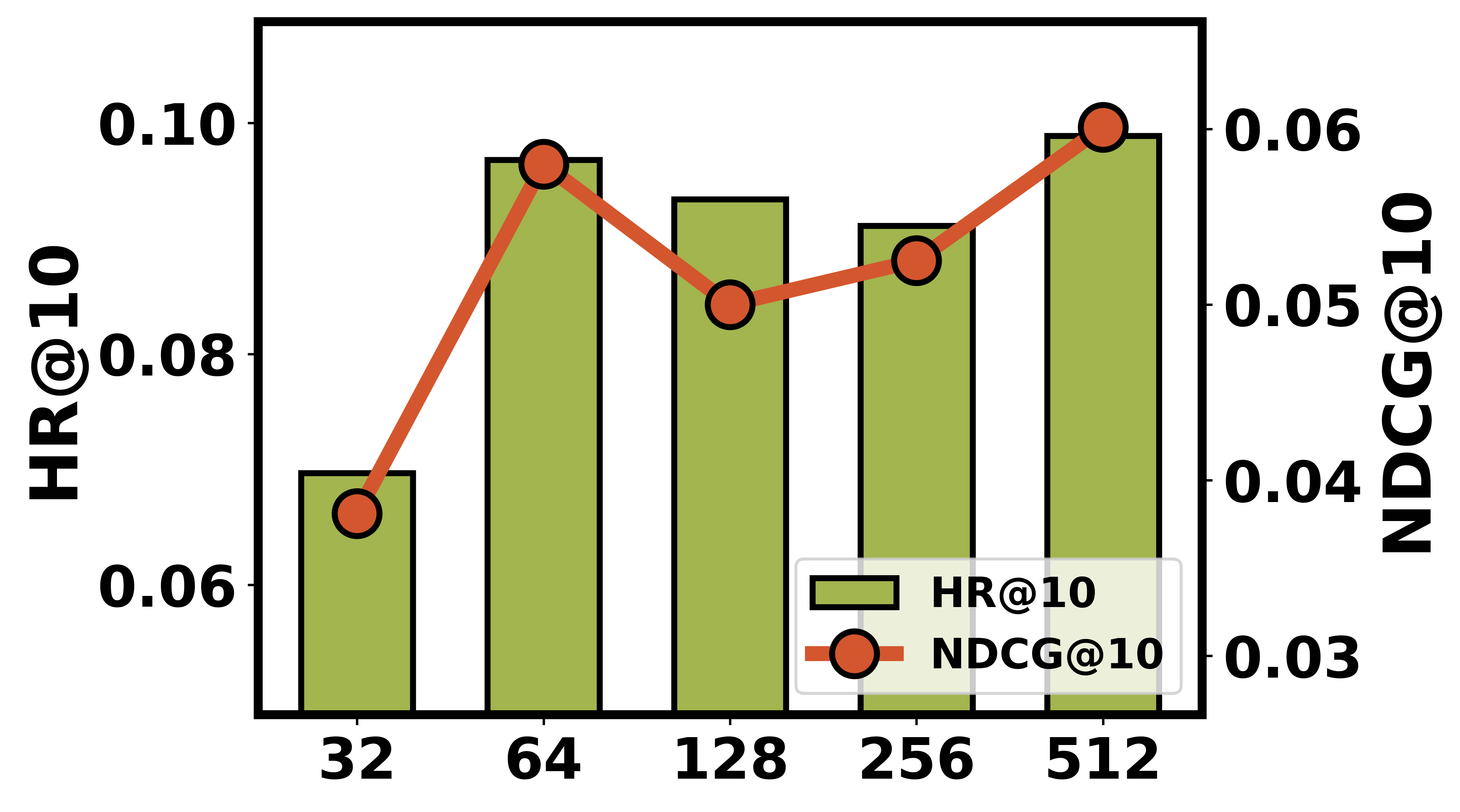}
	}%
 
 	\subfloat[Impact of $\gamma$ on Toys \& Games]{		
		\includegraphics[scale=0.25]{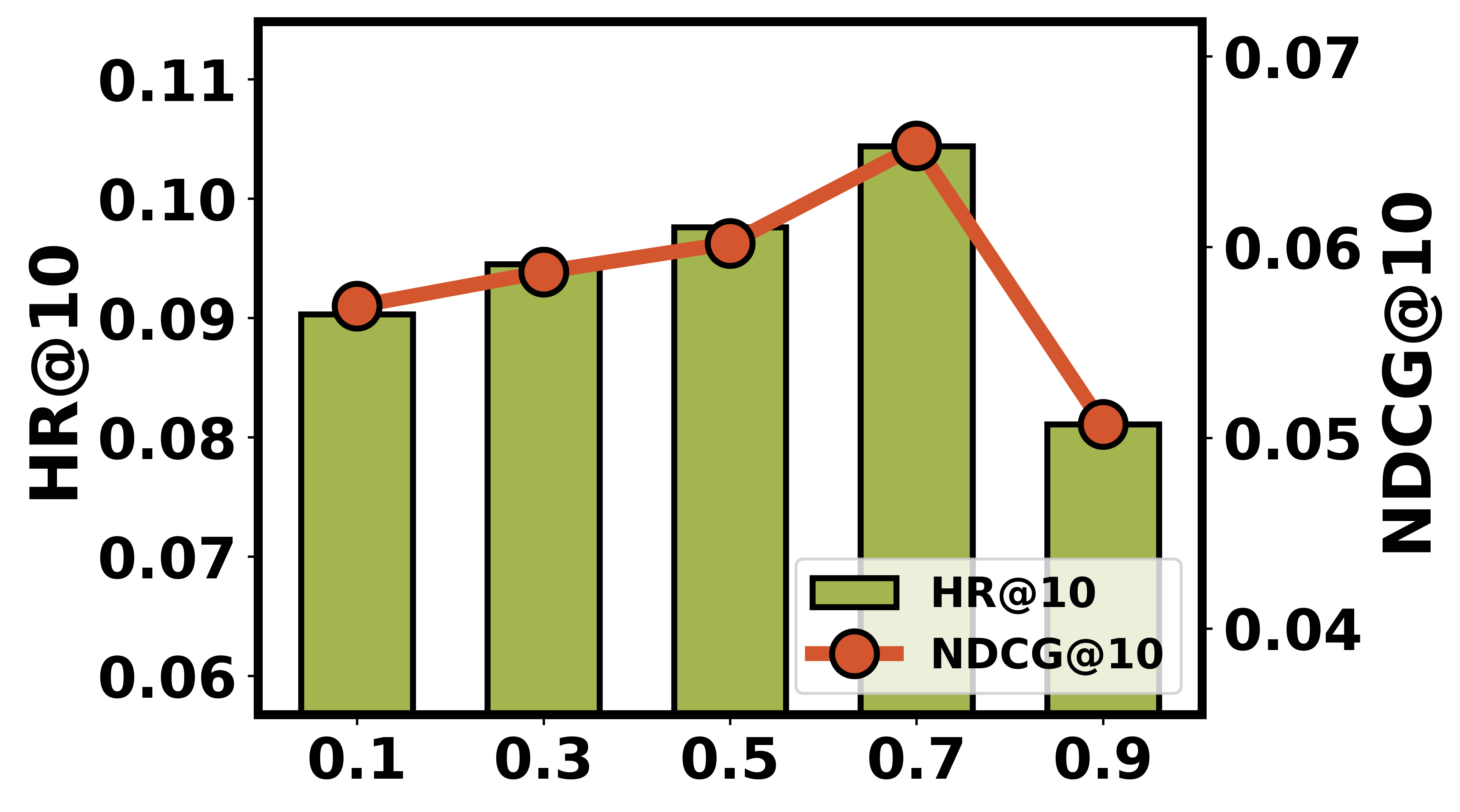}
	}%
 	\subfloat[Impact of $\alpha$ on Toys \& Games]{
		\includegraphics[scale=0.25]{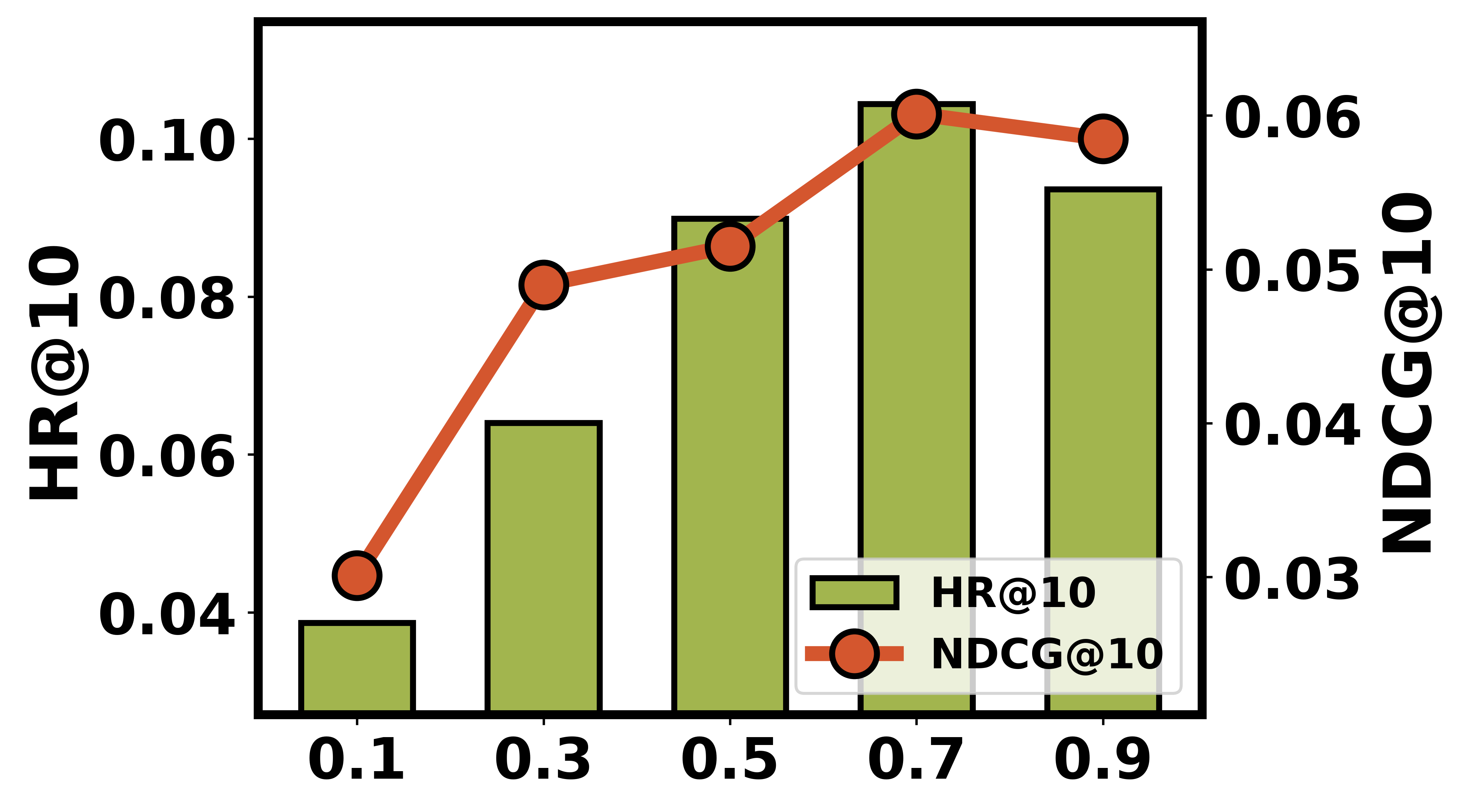}
	}%
	\subfloat[Impact of $\beta$ on Toys \& Games]{		
		\includegraphics[scale=0.25]{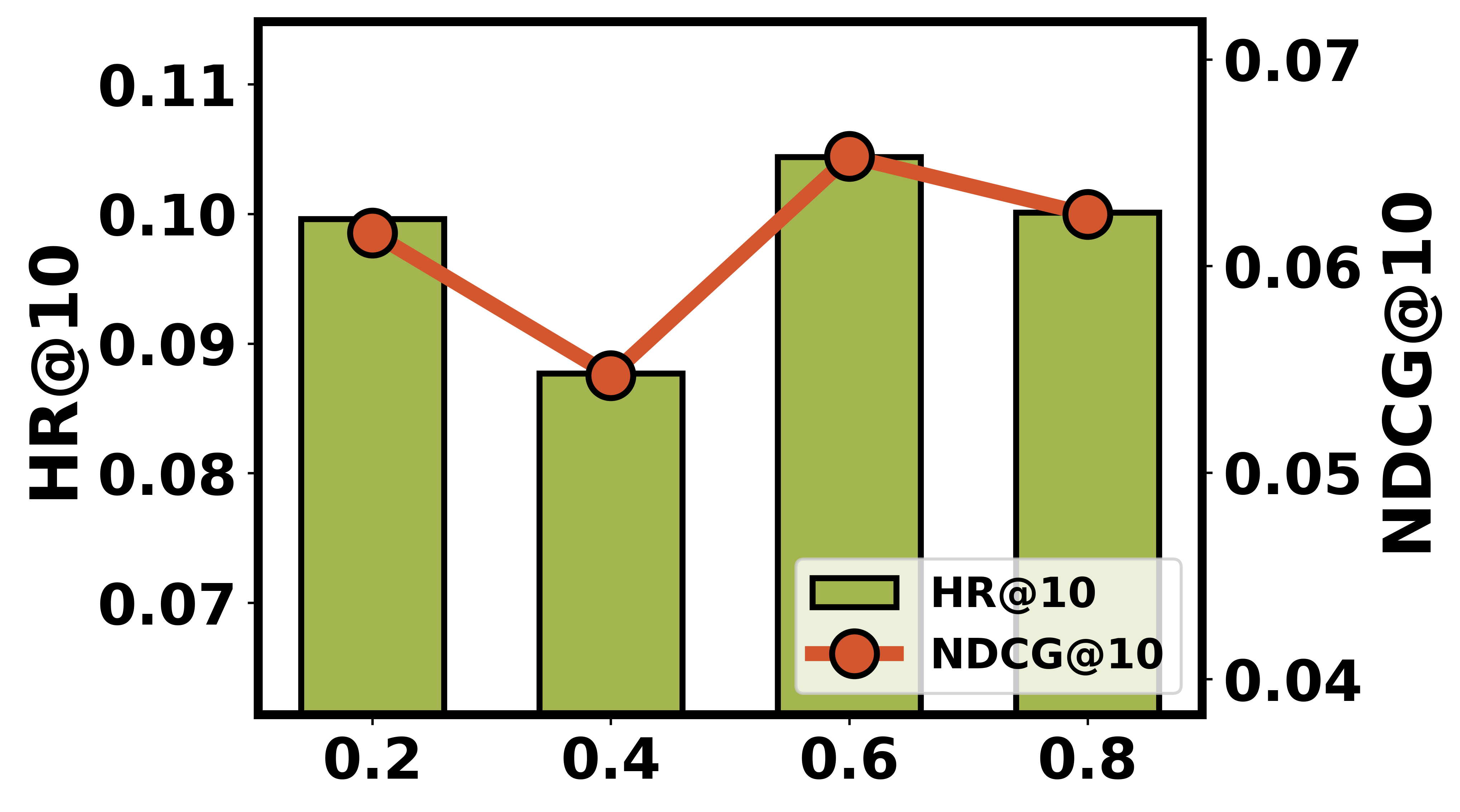}
	}%
 	\subfloat[Impact of $B$ on Toys \& Games]{
		\includegraphics[scale=0.25]{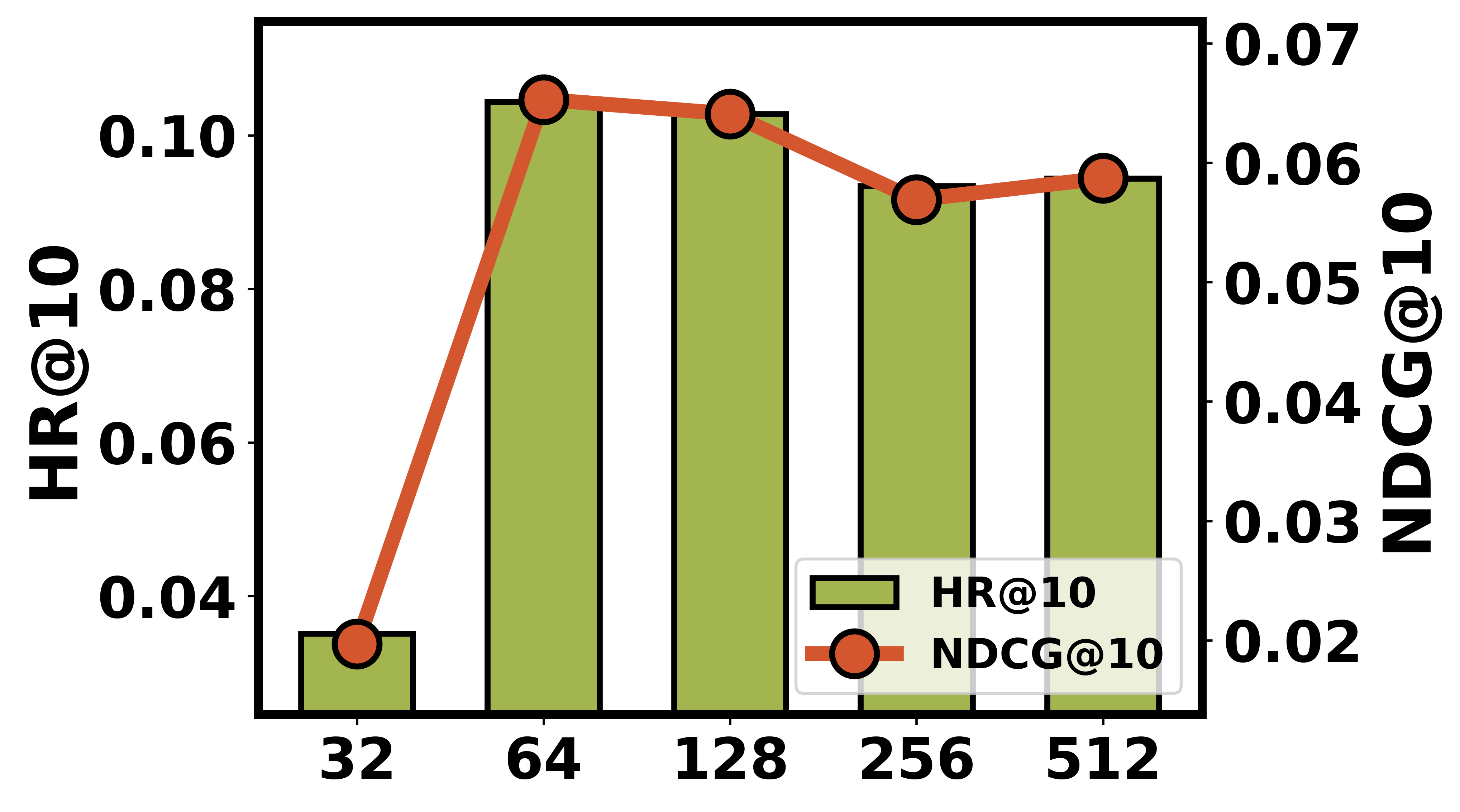}
	}%
	\centering
	\caption{Sensitivity test of hyper-parameters on the Beauty and Toys \& Games datasets.}
	\label{Hyper}
\end{figure*}
\subsection{Effectiveness of Handling Noisy Data}
We benchmark FreqRec under the noise-resilience ability following the protocol of previous work \cite{liu2025preference}. Instead of the standard supervised paradigm, we train the model on the aggregate interaction logs from three Amazon categories, {\tt Automotive}, {\tt CDs \& Vinyl}, and {\tt Grocery \& Gourmet Food}, and then evaluate it on each category independently, with no fine-tuning on the target domain. This evaluation setting requires the model to distinguish meaningful behavioral signals from heterogeneous and domain-specific noise in the training data. 

Table \ref{cross_domain} presents HR@10 and NDCG@10 for each target domain, comparing FreqRec against Frequency-based baselines: FMLPRec, BSARec. FreqRec achieves consistently superior results across all three target categories. Align with the observations in Table \ref{abla}, removing either the $\mathrm{GSA}$ or the $\mathrm{LSR}$ module degrades performance in every domain, confirming that cohort-level denoising and user-level refinement are complementary. Eliminating both frequency-driven modules produces the most pronounced degradation, demonstrating their synergistic effect: the Inter-module suppresses pervasive noise patterns, while the intra-module restores high-frequency, user-specific signals. These results confirm that combining dual-path spectral modules with a frequency-domain consistency loss yields both robust noise suppression and strong generalization.

\subsection{Performance \textit{w.r.t} Data Sparsity}
To assess the impact of data sparsity on model performance,  we focus on users with below-average sequence lengths and group them by interaction count (e.g., 5-6, 7-8) based on their interaction numbers, and evaluate our FreqRec against leading baselines (i.e., SASRec, FMLPRec, BSARec) on {\tt Sport \& Outdoors} and {\tt Beauty} datasets. As shown in Table \ref{sparse}, all models suffer performance drops on sparser sequences. However, FreqRec consistently demonstrates greater stability than the other models. This can be attributed to our approach, which cohort-level aggregation with user-specific refinement, compensating for the information sparsity inherent in short user histories.

\begin{table}
\tiny
\centering
\fontsize{8pt}{8pt} \selectfont 
\renewcommand{\arraystretch}{0.7} 
\setlength{\tabcolsep}{1.8pt} 
\begin{tabular}{@{}c|cccc|cccc@{}}
\toprule
 & \multicolumn{4}{c|}{\textbf{Sports \& Outdoors}} & \multicolumn{4}{c}{\textbf{Beauty}} \\ \cmidrule(l){2-9} 
 & \multicolumn{2}{c|}{[5,6]/51.41\%} & \multicolumn{2}{c|}{[7,8]/20.57\%} & \multicolumn{2}{c|}{[5,6]/50.90\%} & \multicolumn{2}{c}{[7,8]/20.08\%} \\\cmidrule(l){2-9}
\multirow{-3}{*}{Model} & H@5 & \multicolumn{1}{c|}{N@5} & H@5 & N@5 & H@5 & \multicolumn{1}{c|}{N@5} & H@5 & N@5 \\ \midrule
SASRec & 0.0062 & 0.0044 & 0.0041 & 0.0028 & 0.0202 & 0.0135 & 0.0209 & 0.0131 \\
FMLPRec & 0.0061 & 0.0044 & 0.0046 & 0.0029 & 0.0254 & 0.0172 & 0.0080 & 0.0047 \\
BSARec & 0.0170 & 0.0121 & 0.0106 & 0.0093 & 0.0300 & 0.0218 & 0.0213 & 0.0149 \\ \midrule
FreqRec & \textbf{0.0190} & \textbf{0.0135} & \textbf{0.0123} & \textbf{0.0101} & \textbf{0.0325} & \textbf{0.0232} & \textbf{0.0239} & \textbf{0.0163} \\  \bottomrule
\end{tabular}
\caption{Performance of FreqRec against baselines under different sparse interaction degrees.}
\label{sparse}
\end{table}

\subsection{Sensitivity Test on Hyper-Parameters}
Figure \ref{Hyper} presents the hyper-parameter sensitivity analysis of FreqRec, reporting HR@10 and NDCG@10 performance on {\tt Beauty} and {\tt Toys \& Games} datasets.
\begin{itemize}[leftmargin=*]
    \item \textbf{Sensitivity to GSA/LSR learner feature weight $\gamma$ :} In the sub Figure \ref{Hyper} (a, c), performance peaks at $\gamma$ = 0.7 for both datasets, indicating the need for balanced feature constraints between learners.
    \item \textbf{Sensitivity to Self-Attention and FreqNet $\alpha$ :} In the sub Figure \ref{Hyper} (b, f), performance peaks at $\alpha$ = 0.7, showing excessive noise filtering can harm performance.
    \item \textbf{Sensitivity to loss balance parameter $\beta$ :} In the sub Figure \ref{Hyper} (c, g), optimal performance occurs at $\beta$ = 0.6, with performance trends varying across datasets.
    \item \textbf{Sensitivity to Batch Size $B$ :} In the sub Figure \ref{Hyper} (d, h), smaller batches ($B$ = 32) negatively impact performance. Best results for Beauty are at $B$ = 64 or 512, while Toys \& Games peaks at $B$ = 64, highlighting the role of batch size in noise filtering and performance.
\end{itemize}


\section{Conclusion}
In this paper, we present FreqRec, a Frequency-Enhanced Dual-Path Network for sequential recommendation that jointly models inter-session and intra-session dynamics via a learnable, complex-valued Fourier transform. By coupling a Global Spectral Aggregator for cohort-level rhythms with a Local Spectral Refiner for user-specific signals, FreqRec effectively suppresses noise and restores high-frequency behavioral patterns that standard attention mechanisms smooth out. We further introduce a frequency-domain consistency loss, which explicitly aligns the model's predicted spectral coefficients with ground-truth signatures, thereby bridging the gap left by conventional time-domain objectives. Extensive experiments on three public benchmarks show that FreqRec delivers state-of-the-art performance, while maintaining robustness under data sparsity and noisy logs scenarios. 

\section*{Acknowledgements}
We would like to thank the anonymous reviewers for their valuable discussion and constructive feedback. This work is supported by National Natural Science Foundation of China (U22B2061), National Key R\&D Program of China (2022YFB4300603), and Natural Science Foundation of Sichuan, China (2024NSFSC0496).

\bibliography{aaai2026}

\end{document}